\documentclass[12pt]{article}
\usepackage[english]{babel}
\usepackage{graphicx}
\usepackage{epsfig}
\usepackage{amssymb}
\newcommand{\bea}{\begin{eqnarray}}
\newcommand{\eea}{\end{eqnarray}}
\newcommand{\be}{\begin{equation}}
\newcommand{\ee}{\end{equation}}

\newcommand{\nn}{\nonumber}
\begin{document}
\begin{Large}
\begin{center}
{\bf  Signals of the degenerate BESS model at the LHC}
\end{center}
\end{Large}
\vskip 1.0cm
\centerline {R. Casalbuoni$\;^{a,b)}$,
S. De Curtis$\;^{b)}$, M. Redi$\;^{a,b)}$}
\vskip 1.0cm
\centerline{
$a)$ {\it Dipartimento di Fisica, Univ. di Firenze, I-50125
Firenze, Italia.}}
\centerline{
$b)$ {\it I.N.F.N., Sezione di Firenze, I-50125 Firenze,
Italia.}}

\begin{abstract}
\noindent
We discuss the possible signals of the degenerate BESS model at
the LHC. This model describes a strongly interacting scenario
responsible of the spontaneous breaking of the electroweak
symmetry. It predicts two triplets of extra gauge bosons which
are almost degenerate in mass. Due to this feature, the model
has the property of decoupling and therefore, at low energies
(below or of the order of $100~GeV$) it is nearly
indistinguishable from the Standard Model. However the new
resonances, both neutral and charged, should give quite
spectacular signals at the LHC, where the c.o.m. energy will
allow to produce these gauge bosons directly.
\end{abstract}
\newpage
\section{Introduction}
\begin{figure}
\begin{center}
\psfig{file=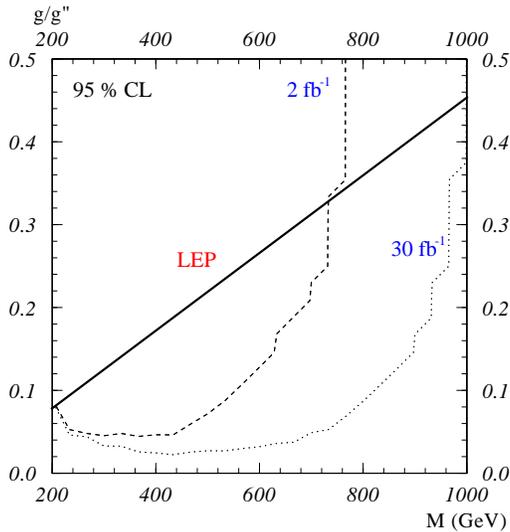,height=7 truecm}
\caption{\it 95\%CL bounds on the D-BESS parameter space
($M$, $g/g^{\prime\prime}$)
 from LEP1/SLC (continuous line) and
Tevatron upgrade at $\sqrt{s}=2~TeV$ with $L=2~fb^{-1}$(dashed
line) and $L=30~fb^{-1}$ (dotted line) assuming no deviation
with respect to the SM in the inclusive cross section
$p\bar{p}\to
\mu\nu_\mu$. Both the statistical error and a systematic error of 5\% on the
cross section have been considered. The allowed regions are
under the curves.}
\label{boundslep}
\end{center}
\end{figure}

As it is well known the Standard Model (SM) is not fully
satisfactory on a theoretical basis, unless supersymmetry is
present to lower the degree of divergence of the scalar sector,
avoiding in this way the hierarchy problem. On the other hand,
if we simply remove the Higgs from the SM, the theory becomes
mathematically incomplete in the sense that it looses the
remarkable property of renormalizability. Moreover, if we think
to a no Higgs scenario, SM violates the unitarity limit in the
scattering of longitudinal $W$ bosons around $1.7~TeV$
\cite{lee}: as a consequence it is widely believed that signals
of the Higgs sector should manifest themselves below this scale
and then will be apparent at the LHC where this energy scale can
be probed. The BESS model \cite{bess} (BESS stands for Breaking
Electroweak Symmetry Strongly) describes a physical scenario
beyond the SM where the spontaneous breaking of the electroweak
symmetry, necessary to give mass to the elementary particles in
the SM, is driven by the dynamics of new strongly interacting
fields around a mass scale of $1~TeV$. The fundamental feature
of the model, in the degenerate version here considered (D-BESS)
\cite{bess1}, consists in the decoupling property of the new
sector of strong interactions at low energies ($Z$ mass). In
fact, since the validity of the SM has been confirmed in the
last years at the impressive level of a few per mill accuracy,
only new physics which very smoothly modifies the predictions of
the SM at the currently accessible energies is still
conceivable. This is actually the problem of the ordinary
technicolor models \cite{technicolor}, which predict, at least
in the simplest QCD rescaled version, rather big corrections to
the LEP observables and then are nowadays experimentally ruled
out. The easiest way to obtain  small deviations from the SM
predictions is to have a new sector which naturally decouples.
In fact in this case, the corrections at the $Z$ energy are
power suppressed in $M_Z/\Lambda_{np}$, where $\Lambda_{np}$ is
the mass scale of the new interactions. The failure of the
technicolor scenario can actually be seen as a consequence of
the fact that $\Lambda_{np}\sim 250~GeV$ so that
$M_Z/\Lambda_{np}$ is not small.\\ Very sketchy, we now review
the model whose details can be found in \cite{bess1}. If the
role of the Higgs boson is taken by new strongly interacting
fields, beside the Goldstone bosons, which will be absorbed by
the ordinary gauge bosons, it is reasonable to assume that also
vector particles will appear under the form of bound states. The
model describes the low energy theory of these new vector states
introducing them in the formalism of the non linear $\sigma$
model as gauge bosons of a local hidden symmetry
$H=SU(2)_L\otimes SU(2)_R$ \cite{nonlinear1} \cite{hiddengauge}.
 This picture has
been inspired by the low energy QCD where the $\rho$ resonance
can be described as a gauge field \cite{nonlinear2} and in fact
QCD itself can be a testing ground. In order to construct the
lagrangian, we then start from the extended symmetry
$G^\prime=G\otimes H$, where $G$ is the ordinary $SU(2)_L\otimes
SU(2)_R$ of the SM which, once made local with respect to the
electroweak subgroup, gives rise to the standard gauge bosons.
The symmetry $G$ is spontaneously broken to the group
$SU(2)_{custodial}$ to protect the relation
$M_Z=M_W/\cos\theta_W$. This breaking produces nine Goldstone
bosons which in turn disappear from the spectrum of physical
particles being absorbed, through the canonical Higgs mechanism,
by $W$, $Z$ and the new gauge bosons which become massive,
leaving a linear combination massless. The appealing property of
decoupling comes from the existence of an accidental symmetry
$[SU(2)_L
\otimes SU(2)_R]^3$  in whose correspondence the two
triplets of extra-gauge bosons are degenerate in mass.\\ In this
letter we present a simulation of the degenerate BESS model at
the LHC (Large Hadron Collider) considering the Drell-Yan
processes where a pair of high energy leptons is produced in the
hard interaction of the protons. In particular the results have
been obtained considering the CMS detector but similar results
should hold also for the ATLAS experiment. \\ Regarding the
study of the model at the existing accelerators, we refer to the
second paper in ref. \cite{bess1}. At LEP1 we can encode the
virtual effects of the heavy resonances in the $\epsilon$
variables \cite{epsilon}, and obtain bounds on the parameters of
the model. In order to compare with the experimental data,
radiative corrections have to be taken into account. Since the
model is an effective description of a strongly interacting
symmetry breaking sector, one has to introduce an UV cut-off
$\Lambda$. We neglect the new physics loop corrections (this can
be rigorously justified from the point of view of a
renormalizable version of the model \cite{besslineare2}) and
assume for D-BESS the same radiative corrections as in the SM
with $M_H=1~TeV$.
 As a consequence of the decoupling, the model
satisfies the severe limits from LEP1 and SLC (see Figure
\ref{boundslep}) for a wide region of the parameter space
without any fine tuning and even for the choice $M_H=1~TeV$, a
value highly disfavoured by the fit within the SM.
\noindent
Limits can also be obtained from the direct search of $W^\prime$
at the Tevatron. The present bounds are not much more
restrictive than those coming from LEP1 (see \cite{tevatron})
but, waiting for the forthcoming upgrade, we can extrapolate
them as shown in Figure \ref{boundslep}. \\ The direct search of
the  D-BESS resonances at the Tevatron upgrade has been
discussed in \cite{tevatron} where a simulation of the model at
the LHC was presented as well. Only the muon channel of the
Drell-Yan processes was considered and a very rough simulation
of the energy smearing was performed. The improvement of this
work consists in a much more realistic simulation of the
response of the CMS detector and in the analysis of the electron
channel. The CMS detector has a better energy resolution in this
channel which will eventually allow to disentangle the nearly
degenerate resonances, a key feature of this model, in the
neutral channel for some choices of the parameters.
\\ In section 2 we briefly recall the main features of the model
underlining the differences with the ordinary strongly
interacting models and giving the relevant formulas. In section
3 we report the results of the simulation at the LHC both for
the neutral and charged processes in the electron and  muon
channels and we also discuss  the possibility to disentangle the
almost degenerate neutral resonances.

\section{The degenerate BESS model}
\label{section2}
The D-BESS model predicts two new triplets of gauge bosons
($L^\pm,~L_3$), ($R^\pm,~R_3$). In comparison with the SM we
have two more parameters, the common strong gauge coupling
$g^{\prime\prime}$ of the triplets and the mass scale of the new
sector. For this it will be convenient to choose the mass $M$ of
the $R^\pm$ resonances.\\ Concerning the coupling to the
fermions, we assume the same couplings as in the SM, which means
that fermions are singlets of the hidden sector, even if a
direct coupling is still consistent with the symmetry of the
lagrangian. With this choice the new physical particles are
coupled to the fermions only via the mixing with the standard
particles.\\The physical fields in the theory are found
diagonalizing the mass matrix whose eigenvalues are the mass of
the particles. In the charged sector $R^\pm$ are not mixed. The
charged fields $W^\pm$ and $L^\pm$ have masses (the following
formulas are all in the limit $M\to \infty$ and
$g^{\prime\prime}\to
\infty$):
\bea
M_{W^\pm}^2&=&\frac {g^2v^2} 2 \nn \\
M_{L^\pm}^2&=&M^2(1+2(\frac{g}{g^{\prime\prime}})^2)
\eea
$g$ the $SU(2)_L$ coupling of the SM and $v^2=1/(\sqrt{2}
G_F)$.\\ In the neutral sector we have:
\bea
M_Z^2&=&\frac {M_W^2} {c^2_\theta} \nn \\ M^2_{L_3}&=&M^2(1+2
(\frac{g}{g^{\prime\prime}})^2)
\nn
\\M^2_{R_3}&=&M^2(1+2 (\frac{g}{g^{\prime\prime}})^2\tan^2\theta)
\eea
where $\tan\theta=s_\theta/c_\theta=g^\prime/g$ and $g^\prime$
the $U(1)_Y$ gauge coupling and we have neglected terms which
are ${\cal {O}}(1/M^2)$. It is important to notice that the new
resonances are degenerate in mass, except small corrections due
to the fact that $U(1)_Y$ explicitly violates the custodial
symmetry. The splitting between the masses in the neutral
channel (the charged one is not really interesting since $R^\pm$
are not coupled to the fermions) at the lowest order is:
\be
\Delta M=M(1-\tan^2\theta)\left(\frac g {g^{\prime\prime}}\right)^2
\label{deltam}
\ee
The couplings of the gauge bosons to the fermions can easily be
obtained from the SM ones. The charged part of the fermionic
lagrangian is
\begin{equation}
{\cal L}_{charged}=
-\left(a_W W_\mu^-+a_L L_\mu^-\right)J_L^{(+)\mu}+ H.c.
\end{equation}
where, in the $M\to \infty$ and
$g^{\prime\prime}\to
\infty$ limit,
\begin{eqnarray}
a_W&=&\frac {g}{\sqrt{2}}\nonumber\\ a_L&=& -g
\frac{g}{g^{\prime\prime}}\label{9.2}
\end{eqnarray}
and $J_L^{(+)\mu}={\bar \psi}_L \gamma^{\mu}
\tau^+ \psi_L$ with $\tau^+$ the combination
$(\tau_1 + i\tau_2)/2$. Since the $R^{\pm}$ are not mixed, they
are not coupled to the fermions. In  the neutral sector the
couplings of the fermions to the gauge bosons are (the photon is
coupled in the standard way):
\begin{equation}
-\frac{1}{2} \bar\psi [
(v_Z^f+a_Z^f\gamma_5)\gamma_\mu Z^\mu+
(v_{L_3}^f+a_{L_3}^f\gamma_5)\gamma_\mu L_3^\mu+
(v_{R_3}^f+a_{R_3}^f\gamma_5)\gamma_\mu R_3^\mu]\psi
\end{equation}
where $v^f$ and $a^f$ are the vector and the axial-vector
couplings given by
\begin{eqnarray}
v_Z^f=A T^L_3+2 B Q_{em},~~~&&~~a_Z^f=A T^L_3\nonumber \\
v_{L_3}^f=C T^L_3+2 D Q_{em},~~~&&~~a_{L_3}^f=C T^L_3\nonumber
\\
v_{R_3}^f=E T^L_3+2 F Q_{em},~~~&&~~a_{R_3}^f=E T^L_3
\end{eqnarray}
and, again in the limit $M\to\infty$, $g/g^{\prime\prime}\to 0$,
\begin{eqnarray}
&A&= \frac{g}{c_{\theta}} ~~~~B= -g\frac{s^2_\theta}{c_\theta}
\nonumber \\ &C&=-\sqrt{2} g \frac{g}{g^{\prime\prime}}~~~~D= 0\nonumber \\ &E&=
\sqrt{2} \frac{g}{g^{\prime\prime}} \frac{g}{c_\theta} \tan^2 \theta ~~~~F= -E
\end{eqnarray}

The total widths in fermions are:
\bea
\Gamma_{L_3}^{fermions}&=&\frac {2\sqrt{2}G_FM_W^2} {\pi} M_{L_3}
(g/g^{\prime\prime})^2 \nn \\
\Gamma_{R_3}^{fermions}&=&\frac {10\sqrt{2}G_FM_W^2} {3\pi} \frac {s_\theta^4} {c_\theta^4} M_{R_3} (g/g^{\prime\prime})^2 \nn \\
\Gamma_{L^\pm}^{fermions}&=&\frac {2\sqrt{2}G_FM_W^2} {\pi} M_{L^\pm} (g/g^{\prime\prime})^2
\eea
Concerning the decay in light gauge bosons we have:
\bea
\Gamma_{L_3}^{WW}=\frac 1 {48} \Gamma_{L_3}^{fermions} \nn \\
\Gamma_{R_3}^{WW}=\frac 1 {80} \Gamma_{R_3}^{fermions} \nn \\
\Gamma_{L_3}^{ZW}=\frac 1 {48} \Gamma_{L_3}^{fermions}
\eea
The fact that the fermionic widths are predominant is a peculiar
feature of the  D-BESS model. Usual models of strong breaking
like technicolor predict an enhancement of the decay in
longitudinal standard bosons, as it follows from the equivalence
theorem \cite{equivalence}. The difference of the D-BESS model
is related to the absence of a direct coupling between the new
gauge fields and the Goldstone bosons which give mass to $W$ and
$Z$. The decay in longitudinal standard bosons comes only from
the mixing which is suppressed by a factor $g/g^{\prime\prime}$,
then the fermionic decay is predominant due to multiplicity. For
the same reason we have no additional contributions (in the
leading order approximation) to the scattering of longitudinal
electroweak bosons and therefore the model has the same
unitarity limits as the SM.
\section{Degenerate BESS at the LHC}
\begin{figure}
\begin{center}
\begin{tabular}{ll}
\hspace{-1.0cm} \mbox{\psfig{file=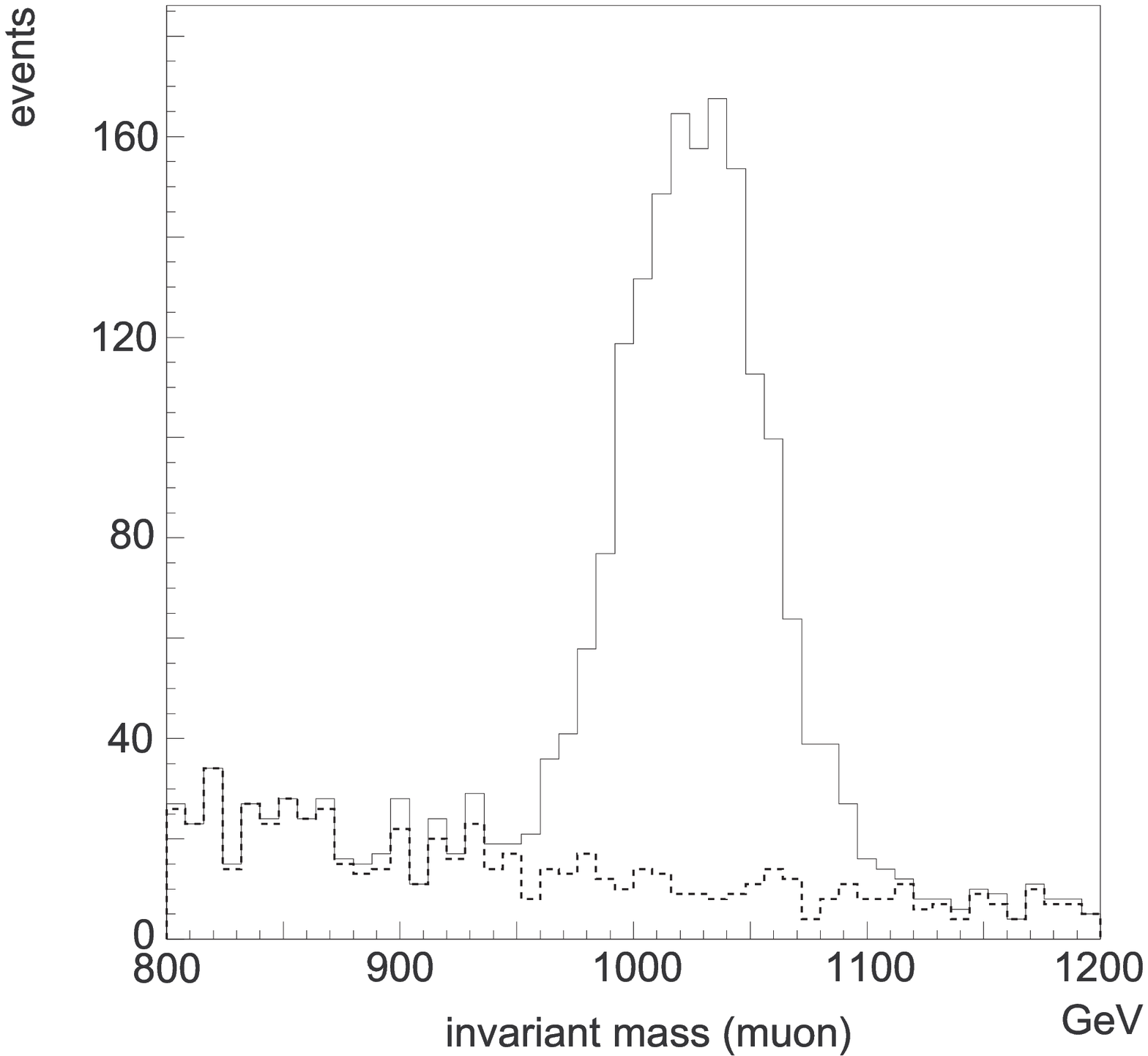,height=7 truecm}} &
\hspace{-.1cm} \mbox{\psfig{file=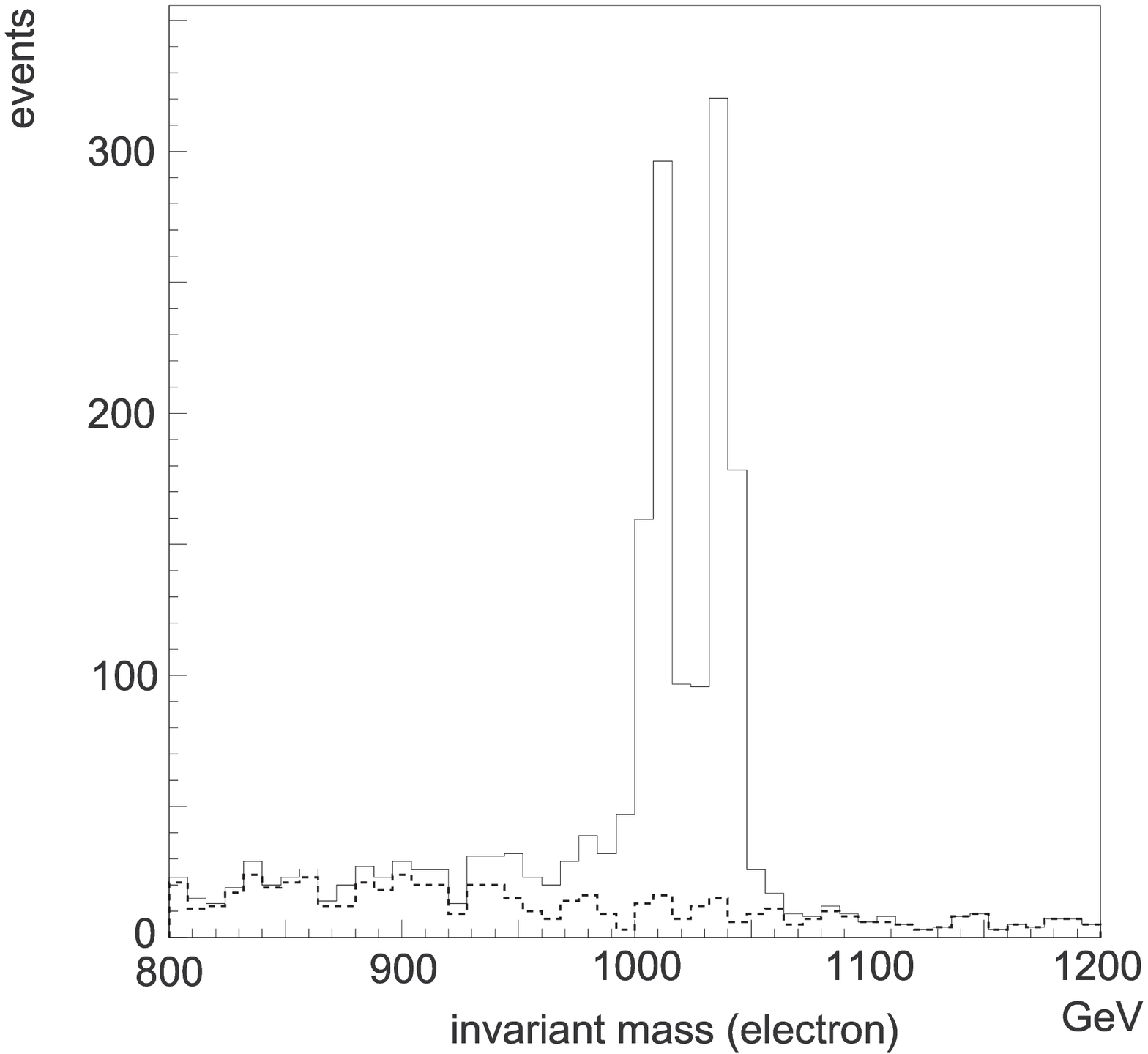,height=7 truecm}}
\end{tabular}
\caption{\it Invariant mass distribution of the events in $pp \to
L_3, R_3, Z, \gamma \to \mu^+ \mu^-$ (left)  and in $pp \to L_3,
R_3, Z, \gamma \to e^+ e^-$ (right) at the LHC, for $M=1000~GeV$
and $g/g^{\prime\prime}=0.2$. The splitting between the
resonances is $29~GeV$. The dashed line is the SM background
while the continuous line represents the D-BESS model prediction
(signal+background). The number of signal and background events
and the applied cuts are given in Table \ref{table2}.}
\label{eneu1002}
\end{center}
\end{figure}

The most suitable machine to study a new sector of strong
interactions is the LHC (Large Hadron Collider), the new
accelerator which will be built to shed light on the still
unknown mechanism of the electroweak symmetry breaking. For a
recent and detailed review of the physics at the LHC we refer to
\cite{proceedings}. The LHC will be able either to discover the
new resonances or to strongly constrain the physical region of
the model.\\ The simulation has been performed considering the
configuration of the LHC with c.o.m. energy $\sqrt{s}=14~TeV$
and an integrated luminosity of $100~fb^{-1}$, corresponding to
one year of run in the so called high luminosity regime
($10^{34}~cm^{-2}s^{-1}$). However in most of the envisaged
configurations the statistical significance of the signal is so
high that even with an integrated luminosity of $10~fb^{-1}$,
deviations from the SM will be observable.\\ The events were
generated using Pythia Montecarlo (version 6.136) \cite{pythia}
and analyzed with CMSJET package \cite{abdoulline} which
performs a simulation of the CMS detector, in particular of the
energy smearing. The Drell-Yan processes, where a pair of
leptons emerges from the $q\bar{q}$ annihilation in a
proton-proton collision, represent the golden plated signature
for the D-BESS model. In fact, as we have already seen in
section \ref{section2}, the decay in ordinary gauge bosons is
 suppressed and, similarly other production processes
like boson scattering are negligible. For each choice of the
parameters of the model (taken inside the region still
unconstrained as shown in Figure \ref{boundslep}) we have
compared the electron and the muon channels. On the theoretical
point of view the processes are identical (except different
radiation effects in the final states) because universality
holds in the BESS model. However the electron channel is
experimentally much more convenient because the CMS detector has
a better energy resolution in this channel than in the muon
channel (di-electron masses in the $TeV$ range can  be hopefully
determined with about 1\% precision \cite{denegri}). The energy
resolution affects dramatically the shape of the resonances
since we are dealing with very narrow resonances: the typical
width for a $1~TeV$ boson is less than $1~GeV$ (see Tables
\ref{table1} and \ref{table2}).\\ In the charged channel we have
considered the transverse mass distributions, for the inclusive
process $pp \to L^\pm,~W^\pm
\to e \nu_e,~\mu \nu_\mu$ comparing them with the SM background.
Since $R^\pm$ are completely decoupled from the fermions they do
not contribute to this process (they can  be produced through
the trilinear vertices of the theory but at a very low rate).
Cuts have been applied in order to maximize the statistical
significance of the signal, basically a cut on the low $p_T$
events which remove the huge SM background coming from $W$
production. We have also imposed an isolation cut on the charged
leptons. The relevant background is then represented by standard
Drell-Yan processes with $W$ exchange and this is the only one
that we have implemented.
\begin{figure}
\begin{center}
\begin{tabular}{ll}
\hspace{-1.0cm} \mbox{\psfig{file=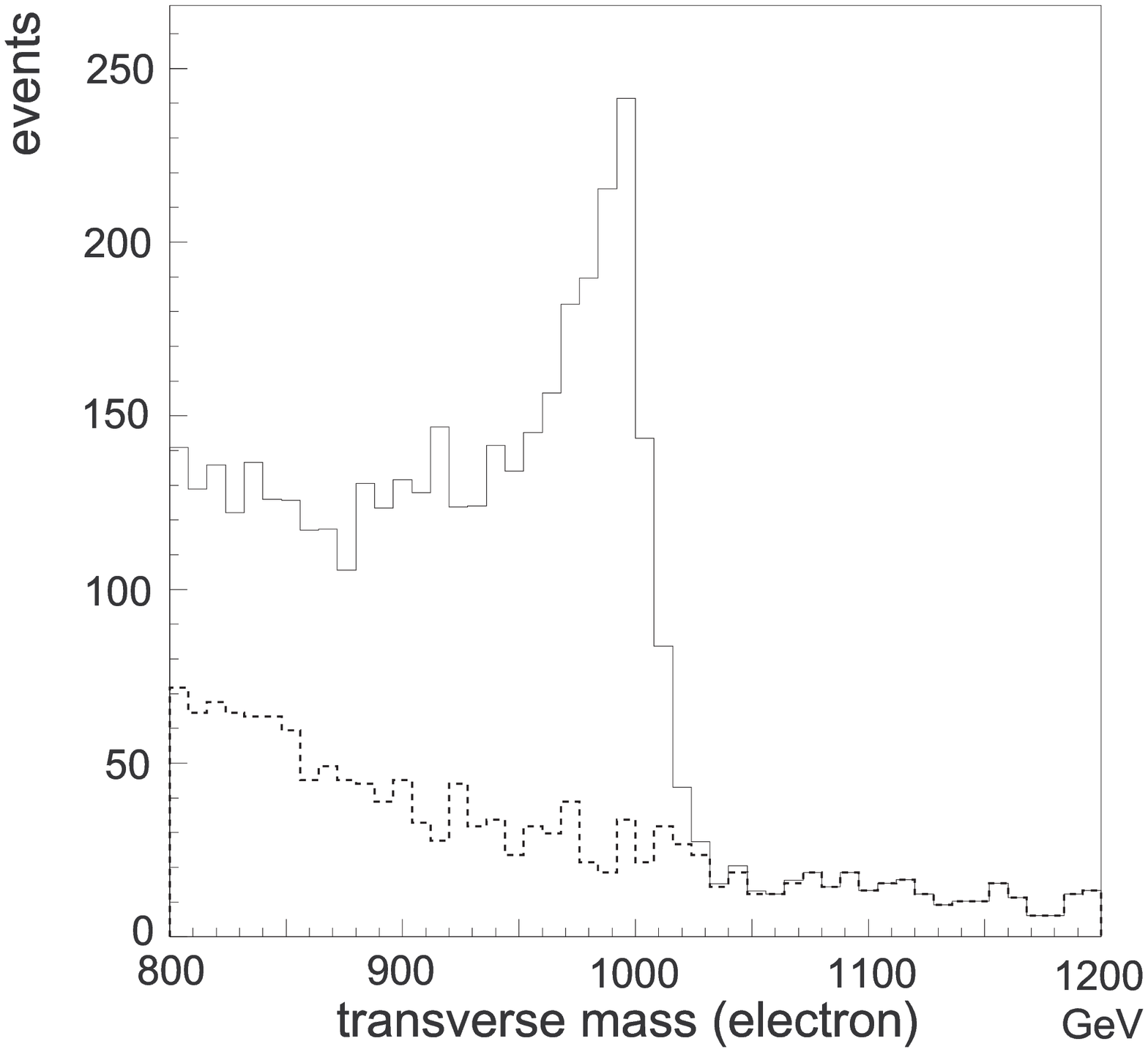,height=7 truecm}} &
\hspace{-.1cm} \mbox{\psfig{file=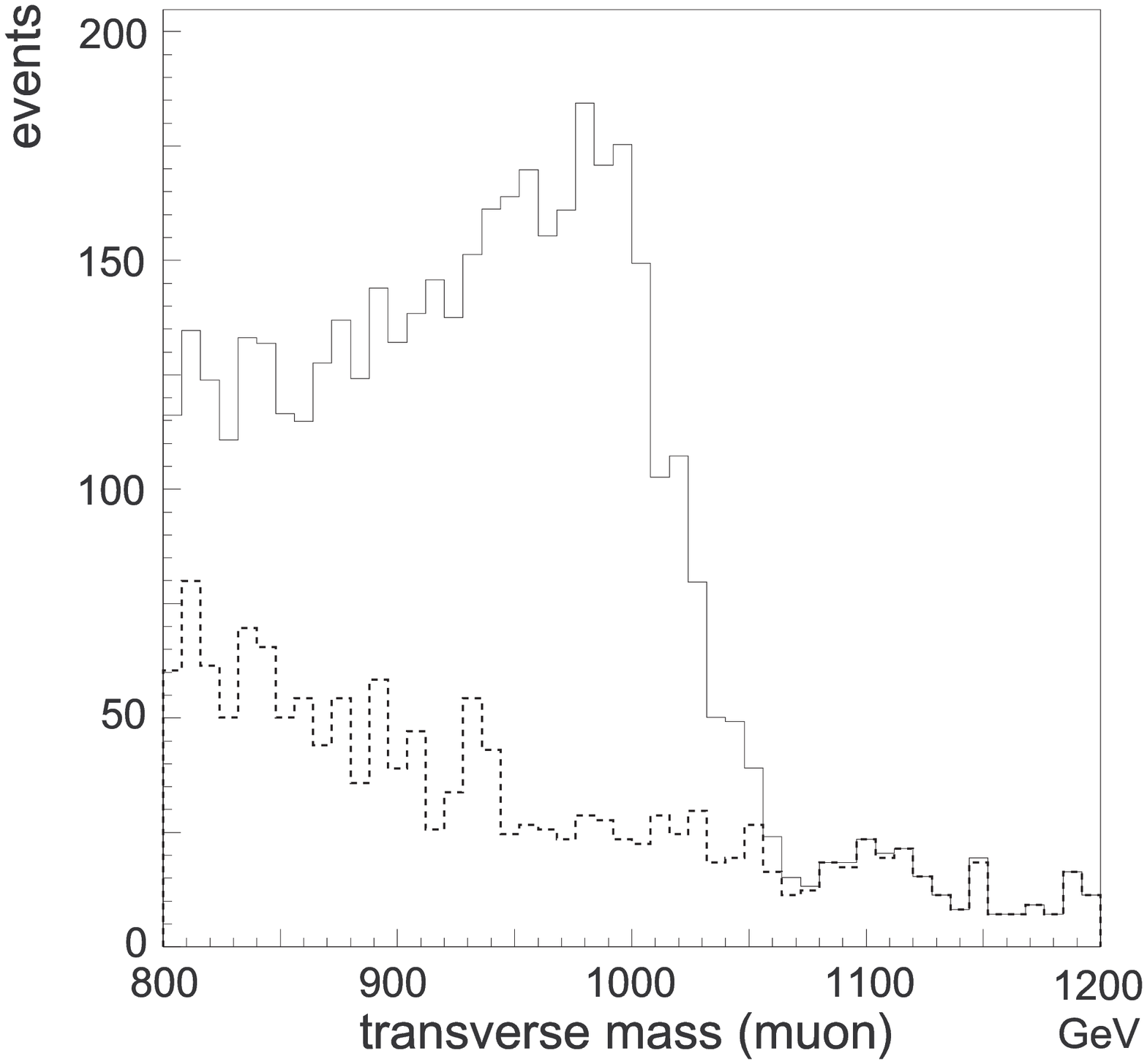,height=7 truecm}}
\end{tabular}
\caption{\it Transverse mass distribution of $pp\to
L^\pm,W^\pm\to e\nu_e$ events (left) and $pp\to L^\pm,W^\pm\to
\mu\nu_\mu$ (right) at the LHC with $L=100~fb^{-1}$,
for $M=1000~GeV$ and $g/g^{\prime\prime}=0.1$. The dashed line
represents the SM while the continuous one is  D-BESS model
(signal+background). The number of signal and background events
and  the applied cuts are given in Table \ref{table1}.}
\label{mtech1000}
\end{center}
\end{figure}
\begin{figure}
\begin{center}
\begin{tabular}{ll}
\hspace{-1.0cm} \mbox{\psfig{file=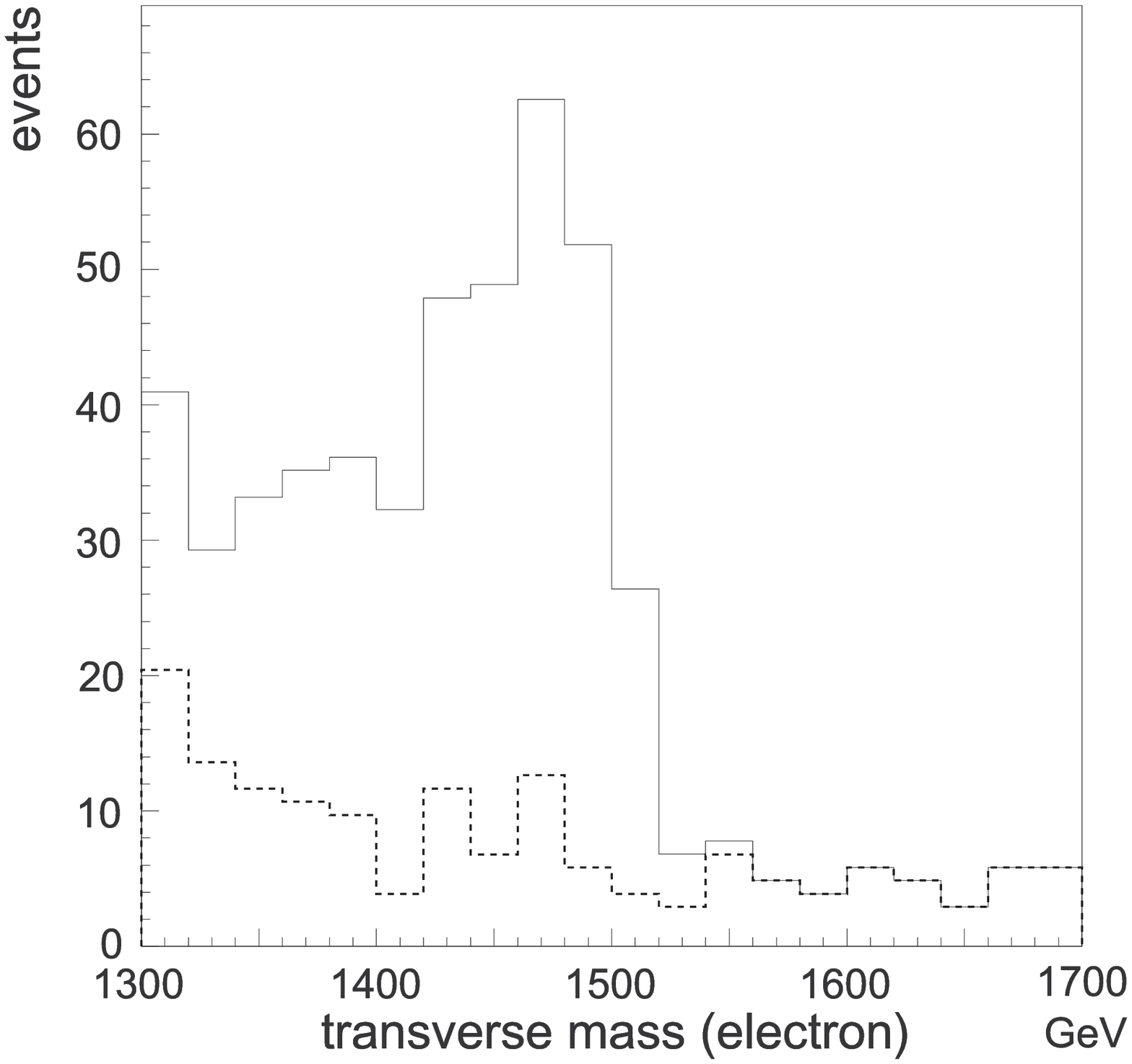,height=7 truecm}} &
\hspace{-.1cm} \mbox{\psfig{file=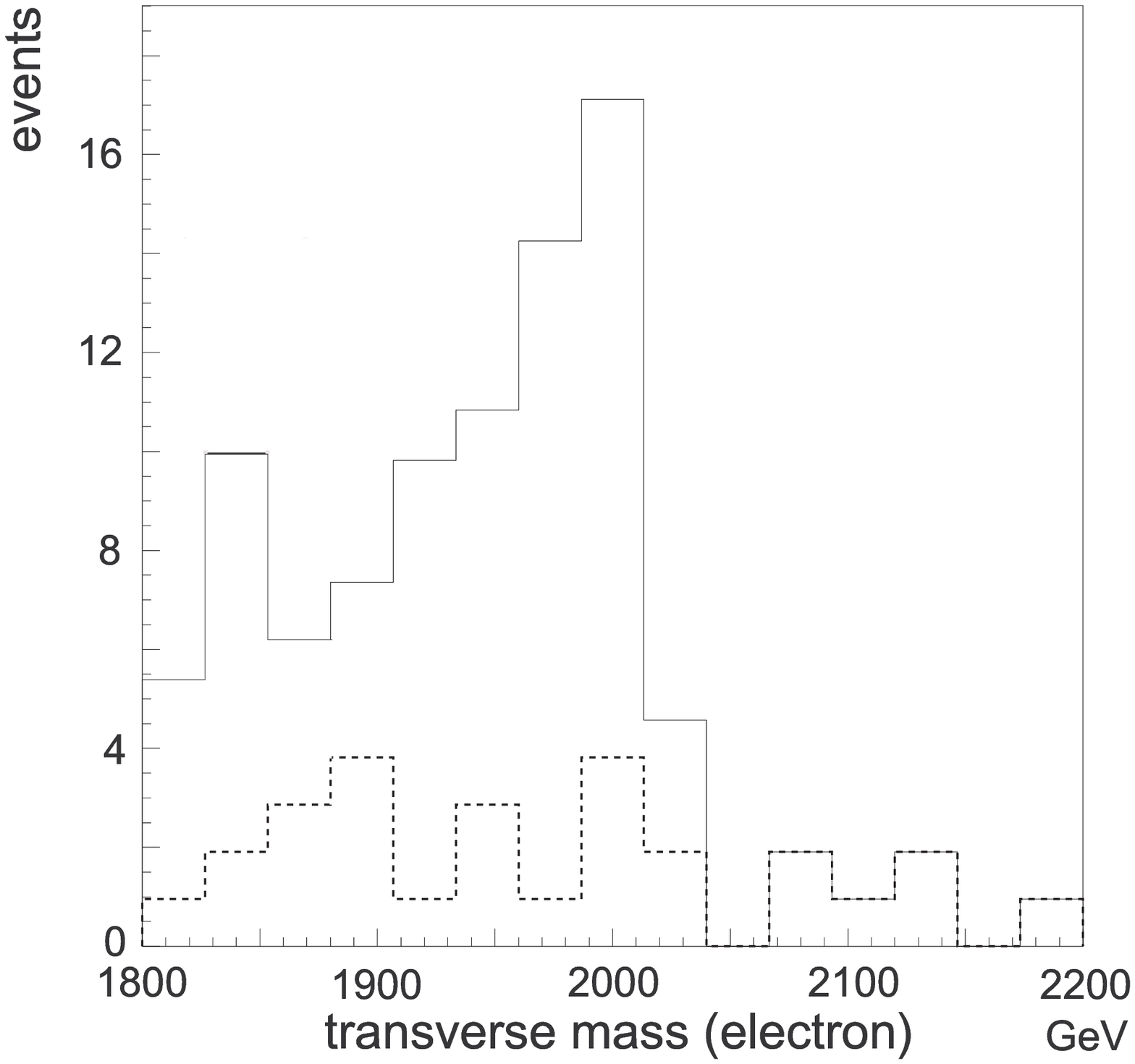,height=7 truecm}}
\end{tabular}
\caption{\it
Transverse mass distribution of $pp\to L^\pm,W^\pm\to e\nu_e$
events at the LHC with $L=100~fb^{-1}$, for $M=1500~GeV$ (left)
and $M=2000~GeV$ (right) and $g/g^{\prime\prime}=0.1$. The
number of signal and background events and the applied cuts are
given in Table \ref{table1}.}
\label{ech1500}
\end{center}
\end{figure}
The distributions of the transverse mass clearly show the
typical jacobian peak around the mass of the new resonances. In
Figure \ref{mtech1000} we compare the transverse mass
distributions in the electron and muon channels both referring
to the same choice of the D-BESS parameters and to the same
applied cuts. We can notice that in the muon channel the
distributions are smoother than in the electron one, due to the
different energy resolution of the CMS detector.
\\
We have analyzed the signature of the D-BESS at the LHC for
several choices of the parameters (see Tables \ref{table1} and
\ref{table2}) chosen inside the physical region and such that
they will not be ruled out by the Tevatron upgrade if no
deviations from the SM are seen, as shown in Figure
\ref{boundslep}. A low mass resonance, for example $M=500~GeV$
with $g/g^{\prime\prime}=0.1$ (still within present bounds)
would give a spectacular signal at the LHC, with a huge number
of events and a very high statistical significance, but we don't
present this case here since it would eventually be discovered
at the Tevatron.
 In Figure
 \ref{ech1500}  we show the
distributions in the electron channel corresponding to
$M=1500,~2000~GeV$ with $g/g^{\prime\prime}=0.1$: the number of
events decreases for increasing mass, but the signal is quite
clear because in this region the SM background is very small.
\\
In the neutral channel we have considered the distribution of
the invariant mass of a pair of leptons. We have applied cuts on
the invariant mass and on the transverse momentum and we have
required that both leptons were isolated in the calorimeters.
These cuts kill all the non Drell-Yan backgrounds like $ZZ$,
$ZW$ or $Z+jets$ or $t\bar{t}$ which in principle can give rise
to decay topologies experimentally indistinguishable from
di-lepton production. Although the number of events is smaller
than in the charged channel, the statistical significance
$S/\sqrt{S+B}$ is very high for low masses of the resonances and
we derive a discovery limit at $2~TeV$ with
$g/g^{\prime\prime}=0.1$. In the neutral channel we have the
chance to detect the nearly degenerate resonances $L_3,~R_3$
which both participate to the process: this would be the most
characteristic signature of the whole model so we especially
concentrate on it. In order to disentangle the double peak
coming from $L_3,~R_3$, a very good energy resolution is
essential,
 so the electron channel looks
the most suitable at CMS. In Figure \ref{eneu1002} we compare
the case $M=1000~GeV$ and $g/g^{\prime\prime}=0.2$ in the
electron and in the muon channel. In the latest, the much wider
experimental width makes it difficult to distinguish the double
peak which is instead visible in the electron channel.
\begin{figure}
\begin{center}
\begin{tabular}{ll}
\hspace{-1.0cm} \mbox{\psfig{file=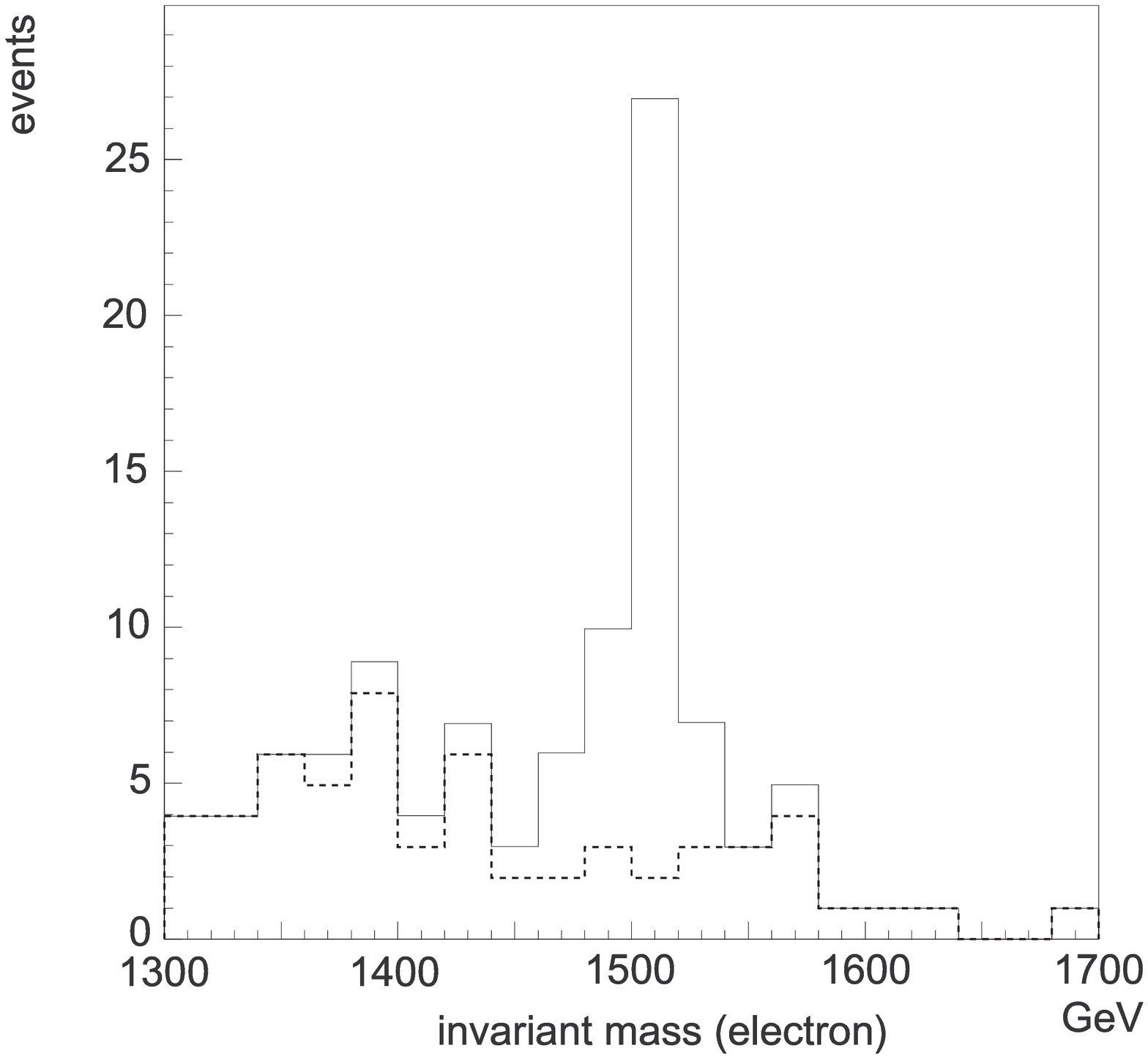,height=7 truecm}} &
\hspace{-.1cm} \mbox{\psfig{file=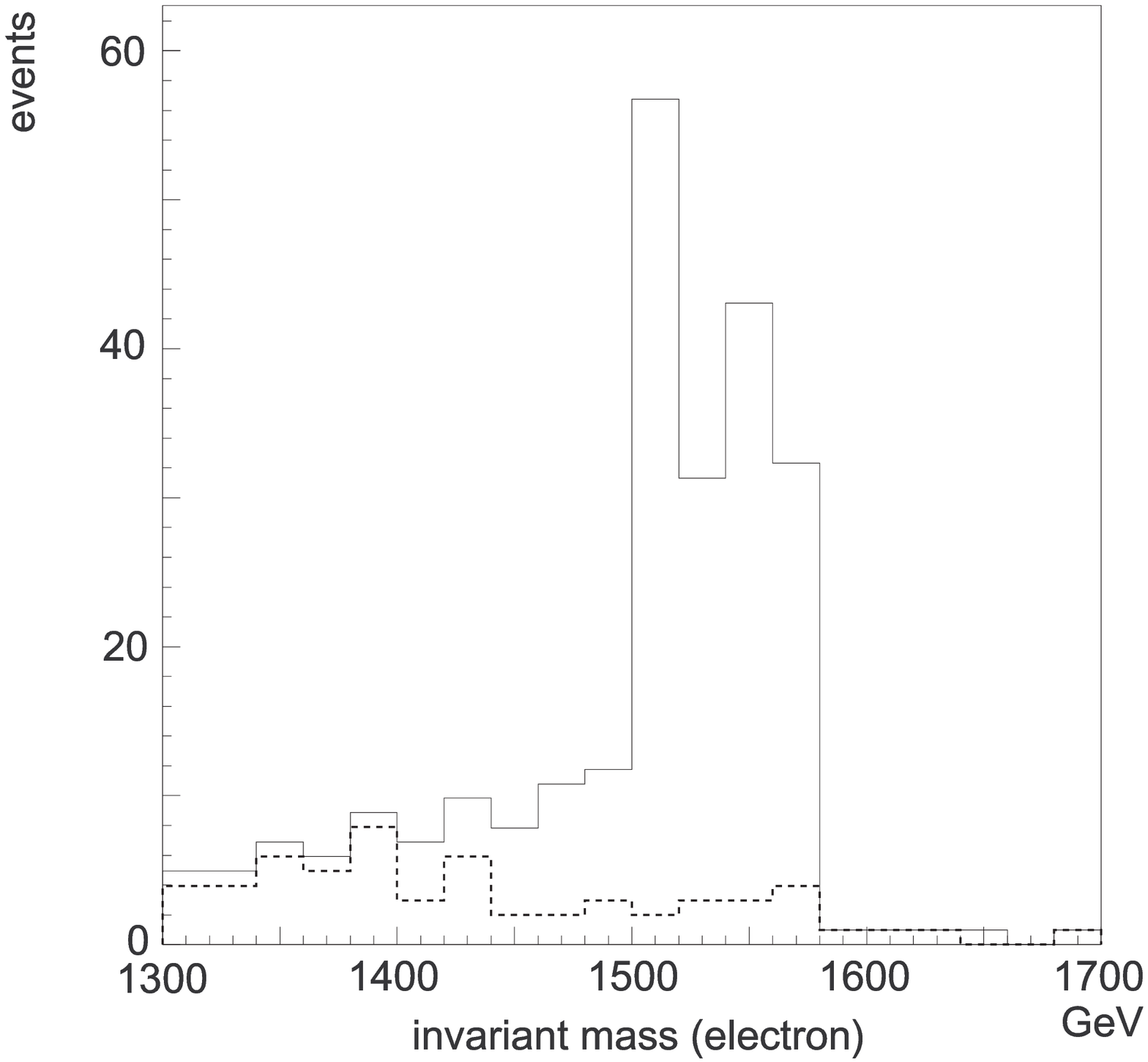,height=7 truecm}}
\end{tabular}
\caption{\it Invariant mass distribution of the events in $pp \to
L_3, R_3, Z, \gamma \to e^+ e^-$ at the LHC, for $M=1500~GeV$
and $g/g^{\prime\prime}=0.1~ {\rm (left)}; ~0.2~ {\rm (right)}$.
The splitting between the resonances is $10.8~GeV$ (left) and
$43~GeV$ (right). The number of signal and background events and
the applied cuts are given in Table \ref{table2}.}
\label{eneu1500}
\end{center}
\end{figure}
\begin{figure}
\begin{center}
\begin{tabular}{ll}
\hspace{-1.0cm} \mbox{\psfig{file=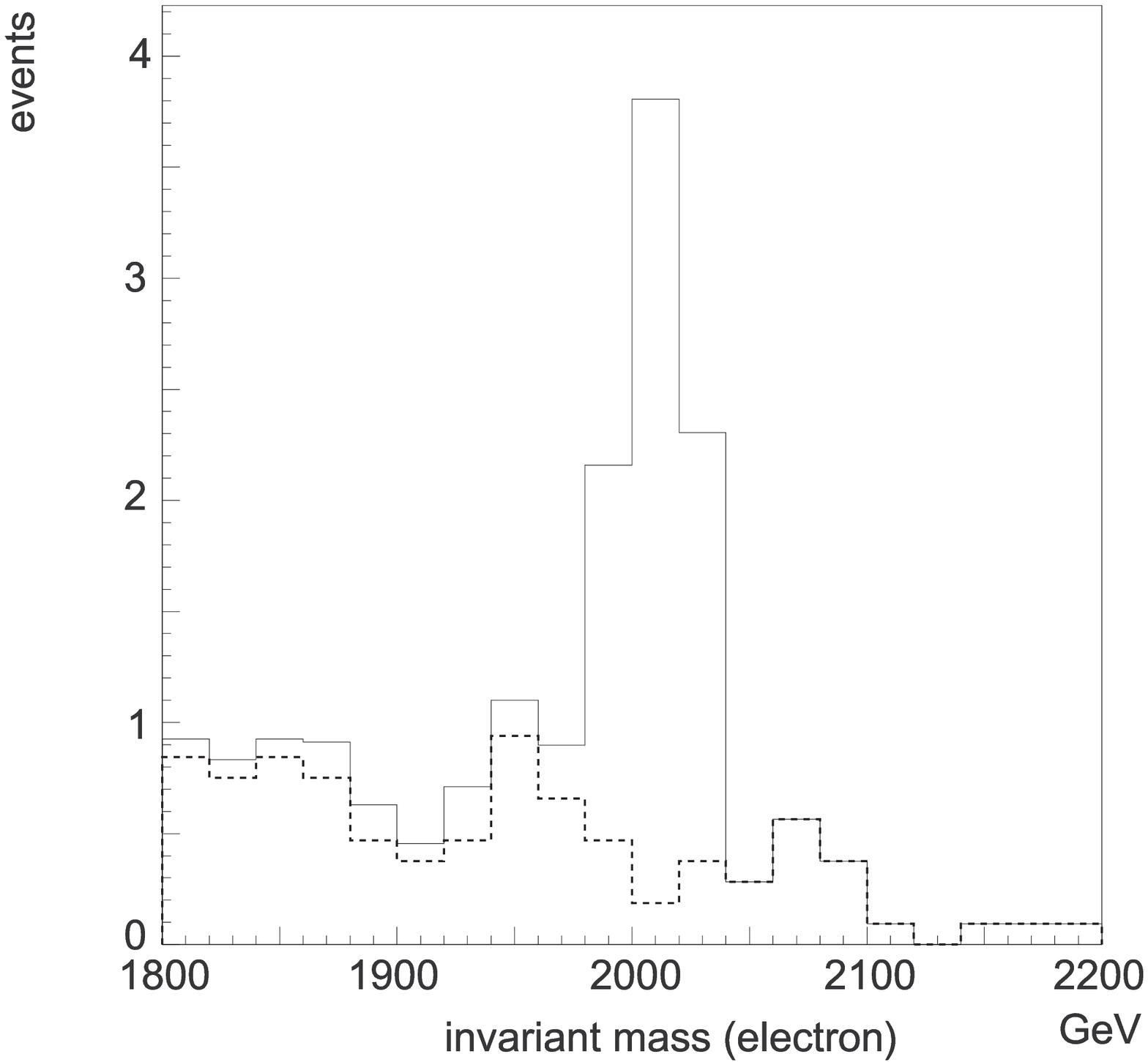,height=7 truecm}} &
\hspace{-.1cm} \mbox{\psfig{file=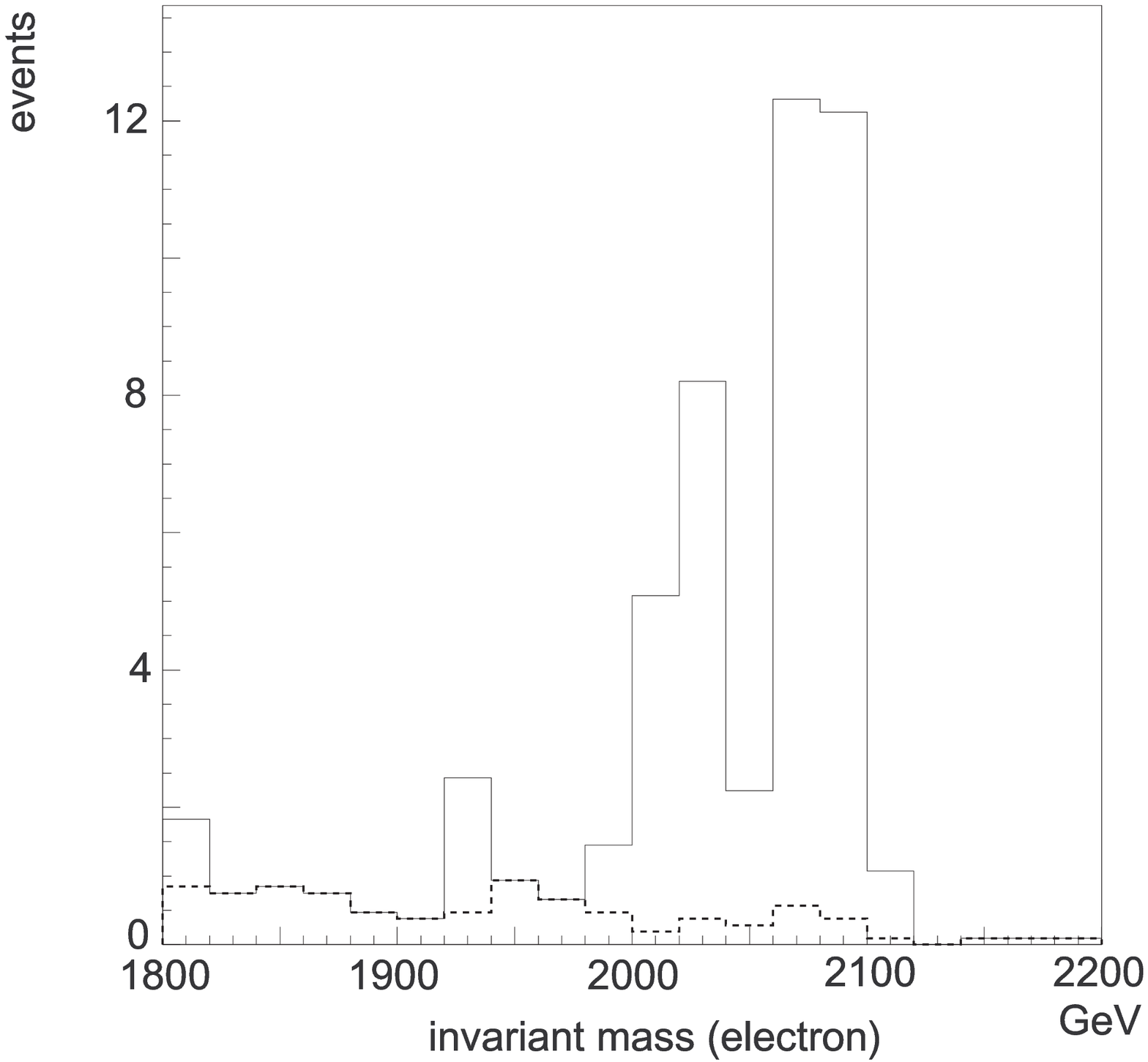,height=7 truecm}}
\end{tabular}
\caption{\it Same as Figure \ref{eneu1500} for $M=2000~GeV$
and $g/g^{\prime\prime}=0.1~ {\rm (left)}; ~0.2~ {\rm (right)}$.
The splitting between the resonances is $14.3~GeV$ (left) and
$57~GeV$ (right).}
\label{eneu2000}
\end{center}
\end{figure}
As in the charged channel we have considered three values of the
mass of the new resonances, $1000~GeV,~1500~GeV$ and $2000~GeV$,
varying $g/g^{\prime\prime}$. The number of signal events at
fixed mass shows the dependence on the parameter
$g^{\prime\prime}$ of the model. In the strong coupling limit
the number of events roughly scales like $1/g^{\prime\prime2}$:
this is once more a consequence of the decoupling property
since, even in the limit $g^{\prime\prime}\to
\infty$, we get back the SM.
On the left-hand side of Figures \ref{eneu1500} and
\ref{eneu2000}, corresponding to $M=1.5,~2~TeV$ with
$g/g^{\prime\prime}=0.1$ the resonances look degenerate in the
invariant mass distribution.
 With the
choices, $M=1,1.5,2~TeV$, and $g/g^{\prime\prime}=0.2$
(right-hand side of Figures \ref{eneu1002}, \ref{eneu1500},
\ref{eneu2000}),
  it is instead possible to disentangle
 the resonances in the electron channel.
 The possibility to distinguish the double peak
depends strongly on $g/g^{\prime\prime}$ and only smoothly on the
mass. In fact, leaving aside the statistical fluctuations, this is
easy to understand: what matters to disentangle the resonances is
the comparison between the energy resolution and the ratio $\Delta
M/M$ which is proportional to $(g/g^{\prime\prime})^2$ (see eq.
(\ref{deltam})) and does not depend on $M$. In the high energy
regime the energy resolution in the di-electron channel can be as
lower as  1\%, leading to a threshold value for
$g/g^{\prime\prime}$ around 0.15. Additionally a higher value of
$g/g^{\prime\prime}$ improves the statistical significance of the
signal.
\\The results of the whole simulation are summarized in Tables \ref{table1}
and \ref{table2} where we also report the cuts used in the
simulation and the widths of the resonances. The number of
events in the muon channel is slightly bigger than in the
electron one and this is mostly due to the isolation cut on the
leptons  we have imposed.
\begin{table}
\begin{center}
\hspace{-1.0cm}
\begin{tabular}{c c c c c c c c c c}
\hline\hline \\
$g/g^{\prime\prime}$ & $M$ & $\Gamma_{L^\pm}$ & $|p_{Tl}^c|$ &
$M_T^c$ & $\# B_e$ & $\# S_e$ & $S_e/\sqrt{S_e+B_e}$& $\#B_\mu$
& $\#S_\mu$
\\
& (GeV) &(GeV) & (GeV) & (GeV) &  &&  \\ \hline \\
0.1 & 1000 & 0.7 & 300 & 800 & 1468 & 2679 & 41.6 & 1529 & 2876
\\ \\
0.1 & 1500 & 1.0 & 500 & 1300 & 154 & 339 & 15.3 & 166 & 422 \\
\\
0.1 & 2000 & 1.4 & 700 & 1800 & 26 & 67 & 6.9 & 31 & 92 \\
\hline\hline
\end{tabular}\\
\caption{\it D-BESS model at the LHC for the process
$pp\to e\nu_e(\mu\nu_\mu)+X$. \#B(\#S) corresponds to the number
of background (signal) events. Also shown are the width of
$L^\pm$ and the applied cuts on the $p_T$ of the lepton and on
the transverse mass.}
\label{table1}
\end{center}
\end{table}
\begin{table}
\hspace{-1.5cm}
\begin{tabular}{c c c c c c c c c c c}
\hline\hline \\
$g/g^{\prime\prime}$ & $M$ & $\Gamma_{L_3}$ & $\Gamma_{R_3}$ &
$|p_{Tl}^c|$ & $m^c_{e^+e^-(\mu^+\mu^-)}$ & $\#B_e$ & $\#S_e$ &
$S_e/\sqrt{S_e+B_e}$ & $\#B_\mu$ & $\#S_\mu$
\\
& (GeV) &(GeV) & (GeV) & (GeV) & (GeV) &&&  \\ \hline \\
0.1 & 1000 & 0.7 & 0.1 & 300 & 800 & 590 & 375 & 12 & 680 &
411\\
\\ 0.2 & 1000 & 2.8 & 0.4 & 300 & 800 & 590 & 1342 & 31 & 680 &
1520\\ \\ 0.1 & 1500 & 1.0 & 0.15 & 500 & 1300 & 58 & 46 & 4.5 &
71 & 69 \\ \\ 0.2 & 1500 & 4.0 & 0.6  & 500 & 1300 & 58 & 189 &
12 & 71 & 247\\ \\ 0.1 & 2000 & 1.4 & 0.2 & 700 & 1800 & 9 & 9 &
2.1 & 12 & 16 \\ \\ 0.2 & 2000 & 5.6 & 0.8  & 700 & 1800 & 9 &
43 & 6.0 & 12 & 52 \\ \\
\hline\hline
\end{tabular}\\
\caption{\it D-BESS model at the LHC for the process $pp\to e^+e^-(\mu^+
\mu^-)+X$. \#B(\#S)
corresponds to the number of background (signal) events. Also
shown are the widths of $L_3$ and $R_3$ and the applied cuts on
the $p_T$ of the leptons and on the invariant mass. }
\label{table2}
\end{table}
\section{Conclusions}
We have shown the possible signals at the LHC of a dynamical
symmetry breaking model called degenerate BESS, which predicts
two triplets of new vector resonances, almost degenerate in
mass. The model has the appealing property of decoupling which
makes it automatically pass all the low energy tests, leaving
the possibility of a strong sector of interactions even at
rather low energies.\\ The neutral channel represents the
clearest signature of the model if it is possible to disentangle
the two resonances. However, the production rate is less
favorable than in the charged one, exactly as it happens in the
SM, since we have considered a minimal coupling to fermions. The
neutral gauge bosons can be detected over the background up to
$M=2000~GeV$, provided $g/g^{\prime\prime}$ is not smaller than
$0.1$. In the charged channel, where only the $L^\pm$ bosons
contribute to the signal, the limit of detection at $2~TeV$ is
reached for $g/g^{\prime\prime}=0.03$, but the experimental
proof of the model requires that both neutral and charged
resonances are discovered. We have compared the electron and
muon channels coming to the conclusion that the first is
experimentally more convenient since the CMS detector has a
better energy resolution for the electrons than the muons in the
high energy range.\\If the new particles will not be discovered
at the LHC, very restrictive limits on the parameter space could
be derived, which, in the light of the unitarity limit of the
model, will essentially close the physical region of the D-BESS
model.
\begin{figure}
\vskip -1cm
\centering
\includegraphics[height=85mm,bb=56 246 543 750]{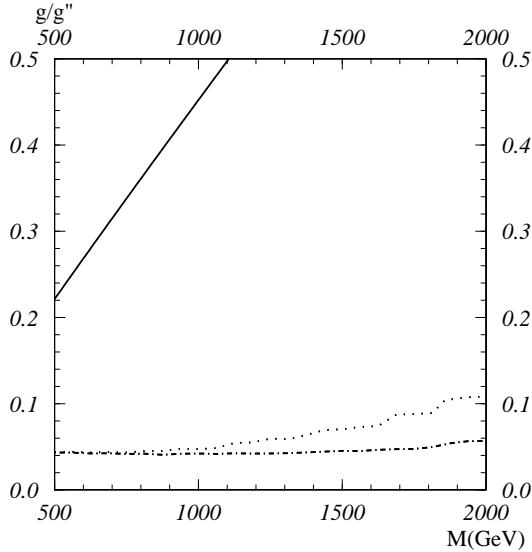}
\caption{\it 95\%CL limits on the parameter space
($M$, $g/g^{\prime\prime}$) of the  D-BESS model at the LHC with
$L=100~fb^{-1}$ (dotted-dashed line) and $L=10~fb^{-1}$ (dotted
line) assuming no deviations are seen from the SM in the cross
section $pp
\to
\mu\nu_\mu$.
Both the statistical error and a systematic error of 5\% have
been taken into account. The continuous line corresponds to the
present LEP1/SLC bound.}
\label{boundslhc}
\end{figure}
In fact, if no deviations from the SM are seen at the LHC, the
bounds shown in Figure \ref{boundslhc} can be drawn.
\eject
\noindent
{\bf{Acknowledgement:}}\\
\noindent
We thank S. Abdulline for allowing  us  to use the
 CMSJET package and for his helpful advice.

\newpage

\newpage


\end{document}